\documentclass[twocolumn,nofootinbib]{revtex4-1}
\usepackage{latexsym,epsfig,amssymb, amsmath,nicefrac}

\usepackage{amsfonts,amsthm} 
\usepackage{bm,bbm}
\usepackage{graphicx}
\usepackage{esint}
\usepackage{color}
\usepackage{braket}

\newcommand{\be}{\begin{equation}}
\newcommand{\ee}{\end{equation}}
\newcommand{\ba}{\begin{eqnarray}}
\newcommand{\ea}{\end{eqnarray}}
\def\bs{\begin{subequations}}
\def\es{\end{subequations}}

\def\Kahler{K\"{a}hler~}

\begin{document}

\title{{\Large Universality classes of inflation}}

\author{Diederik Roest}

\affiliation {Centre for Theoretical Physics, University of Groningen\\ Nijenborgh 4, 9747 AG Groningen, The Netherlands, d.roest@rug.nl}

\begin{abstract}
We investigate all single-field, slow-roll inflationary models whose slow-roll parameters scale as $1/N$ in the limit of a large number of e-folds $N$. We proof that all such models belong to two universality classes, characterised by a single parameter. One class contains small field models like hilltop inflation, while the other class consists of large field models like chaotic inflation. We give the leading expressions for the spectral index and tensor-to-scalar ratio $r$, which are universal for each class, plus subleading corrections for a number of models. This predicts $r$ either to be unobservably small, $r<0.01$, or close to the present observational limit, $r \approx 0.07$.
\end{abstract}

\maketitle

\bigskip

\noindent {\bf Introduction}
The Planck satellite has measured the temperature fluctuation of the cosmic microwave background with unprecedented precision, leading to the following spectral index for primordial fluctuations \cite{Planck}:
 \begin{align}
  n_s & = 0.9603 \pm 0.0073 \,. \label{ns}
 \end{align}
This establishes a percent-level deviation from the Harrison-Zel'dovich scale-invariant spectrum with $n_s =1$; the latter is ruled about at over $5\sigma$. Moreover, Planck has placed a stronger constraint on the ratio between the power spectra of tensor and scalar perturbations: $ r < 0.11$. No evidence has been found for e.g.~non-Gaussianities, isocurvature perturbations or a running spectral index.

Inflation provides a compelling explanation of such perturbations as quantum fluctuations during this phase of exponential expansion. The cosmological observables translate into properties of the inflationary model at the moment of horizon crossing, around 50 to 60 e-folds before the end of inflation. In this paper we will restrict to the simplest case of single-field, slow-roll inflation, consistent with Planck. The Lagrangian of the inflaton field
 \begin{align}
  \mathcal{L} = \sqrt{-g} [ - \tfrac12 (\partial \phi)^2 - V(\phi) ] \,,
 \end{align}
gives rise to the  following cosmological parameters
 \begin{align}
  n_s = 1 + 2 \eta - 6 \epsilon \,, \qquad r = 16 \epsilon \,,
 \end{align}
in terms of the two slow-roll parameters (setting $M_{\rm Pl} = 1$)
 \begin{align}
  \epsilon = \frac12 \left( \frac{V'}{V} \right)^2 \Big\vert_{\phi_*} \,, \qquad \eta = \frac{V''}{V} \Big\vert_{\phi_*} \,.
 \end{align}
In terms of the potential energy, the number of e-folds is
 \begin{align}
  N = \int^{\phi_*}_{\phi_{\rm end}} \frac{V}{V'} \,,
 \end{align}
where the range of inflation runs from horizon crossing at $\phi_*$ to the point $\phi_{\rm end}$ where the slow-roll conditions are violated.

The deviation \eqref{ns} from the scale-invariant spectrum places constraints on different inflationary models; indeed a number of models is now ruled out. Instead of a case-by-case analysis, however, it would be highly desirable to have an organising principle that applies to classes of models. We provide such a principle in this paper. In particular, we will analyse {\it all single-field slow-roll inflationary models that give rise to a spectral index whose deviation from scale invariance scales with $1/N$}. For around 50 to 60 e-folds, this naturally gives rise to percent-level numbers, as requested by Planck. Such models therefore naturally fall in the observationally viable region of cosmologically parameters. 

We will demonstrate that the single assumption of $1/N$ dependence leads to intriguing scaling relations between the slow-roll parameters. It follows that there are only two universality classes of models, one that generically corresponds to small field inflation and one to large field. All single-field slow-roll models asymptote to these universality classes in the limit of large-$N$. Even subleading corrections will be found to satisfy the asymptotic relations. The observational predictions within universality classes are virtually identical. We confront these with the Planck results and derive a generic prediction for the tensor-to-scalar ratio $r$. Related results can be found in \cite{Mukhanov}, where the same universality classes were derived by different means. Similarly, our results complement previous work in which sets of models have been shown to asymptote to examples of these classes, see e.g.~\cite{Westphal, conformal-KL, non-minimal-KL}.

\bigskip

\noindent {\bf Asymptotic slow-roll relations}
Our central assumption is that both slow-roll parameters $\epsilon$ and $\eta$ have an asymptotic power-law dependence on the number of e-folds; in other words both scale as $1/N^p$ for some $p$ at leading order in the limit of large-$N$ (a similar expansion was considered from the effective field theory point of view in \cite{Boyanovsky}). Moreover, given the Planck results, we will assume that $p=1$ for at least one of the two parameters. We will parametrise this dependence as 
 \begin{align}
  \epsilon \simeq \frac{\epsilon_1}{N} \,, \qquad \eta \simeq \frac{\eta_1}{N} \,, \label{expansion}
 \end{align}
where either $\epsilon_1$, $\eta_1$ or both is assumed to be non-vanishing. Throughout the paper, the symbol $\simeq$ means that we are suppressing higher-order terms in $1/N$.

Taking the expansion of the second slow-roll parameter as the starting point, we have the approximate identity
 \begin{align}
   \frac{V''}{V}  \Big\vert_{\phi_*} \simeq \frac{\eta_1}{\int^{\phi_*}_{\phi_{\rm end}} \frac{V}{V'}} \,,
 \end{align}
evaluated at horizon crossing $\phi_*(N)$. Both sides of the above equation have the same $\phi_*$ dependence at leading order, valid for a range of values $\phi_*(N)$  that correspond to large $N$. However, if one is only interested in the leading terms, this equation in fact holds for the entire range of field values: it becomes the functional identity
 \begin{align}
   \frac{V''}{V} \simeq \frac{\eta_1}{\int \frac{V}{V'}} \,. \label{eta-starting-point}
 \end{align}
With this understanding, it is justified to manipulate this equation in order to extract information on the leading order of inflationary parameters. Firstly, we rewrite the above as 
 \begin{align}
  \int \frac{V}{V'} \simeq \frac{\eta_1  V}{V''} \,.
 \end{align}
This equation can be differentiated and multiplied to yield
 \begin{align}
  \frac{(V^{- \eta_1} V')'}{V^{- \eta_1} V'} \simeq \frac{(V''^{-\eta_1})'}{V''^{-\eta_1}} \,.
 \end{align} 
Both sides can be integrated to
 \begin{align}
  \log( V^{-\eta_1} V') \simeq \log ( V''^{- \eta_1} ) + c \,, 
 \end{align}
with an integration constant $c$. Exponentiation then gives
 \begin{align}
  V' \simeq \lambda \left( \frac{V''}{V} \right)^{- \eta_1} \,,
 \end{align}
where the previous integration constant leads to an arbitrary coefficient $\lambda$ between the two sides of this equation. In other words, it should be taken to imply that both sides scale the same at leading order in $N$. It can be rephrased as
 \begin{align}
  \epsilon^{1/2} \eta^{\eta_1}  \simeq \frac{\lambda}{V} \,, \qquad (\eta_1 \neq 0) \,.  \label{eta-rel}
 \end{align}
An analogous analysis for the first slow-roll parameter yields
 \begin{align}
 \epsilon^{2 \epsilon_1}  \simeq  \frac{\lambda}V \,, \qquad (\epsilon_1 \neq 0) \,. \label{eps-rel}
 \end{align}
These asymptotic relations between the slow-roll parameters, valid at large-$N$, will be central in the analysis. Note that these relations are fundamentally different from the relation  $(V / \epsilon)^{1/4} \approx 7 \cdot 10^{16}$ GeV that follows from the COBE measurement of the power spectrum; the latter concerns the actual {\it values} while the asymptotic relations only concern the {\it scaling} behaviour.

Care must be taken in the singular cases where either $\epsilon_1$ or $\eta_1$ vanishes. In the starting points above we have assumed these to be non-vanishing, and strictly speaking the asymptotic relations \eqref{eta-rel} and \eqref{eps-rel} do not apply in these singular limits. Starting with $\epsilon_1 = 0$ and taking the first non-vanishing term to be $\epsilon_p / N^p$ with $p>1$ instead, the analogon of \eqref{eta-starting-point} can be manipulated into 
 \begin{align}
 1 + \frac{4 \epsilon_p^{1/p}}{p} \epsilon^{1-1/p}  \simeq \lambda  V^{-2+2/p}  \,, \qquad (\epsilon_1 = 0) \,. \label{eps-sing-rel}
 \end{align}
This implies that $V$ has to be a constant in the large-$N$ limit: $V \rightarrow V_0$. Note that this behaviour is identical to the $\epsilon_1 \rightarrow 0$ limit of \eqref{eps-rel}.  We have not been able to derive a similar aymptotic relation from the analogon of \eqref{eta-starting-point} for the singular case $\eta_1 = 0$. Instead, taking the first term to be $\eta_p / N^p$ and starting from the weaker relation $\epsilon^p \sim \eta$, one can proof that
 \begin{align}
  \epsilon^{-p+1} \simeq \lambda V^{2p-2} \,, \qquad (\eta_1 = 0) \,.
 \end{align}
Note that this again coincides with the $\eta_1 \rightarrow 0$ limit of \eqref{eta-rel}.

\bigskip

\noindent {\bf Classification}
Based on the Planck results we will assume that at least one of the two leading coefficients is non-vanishing. This leads to three distinct possibilities. 

The first possibility assumes that $\epsilon_1$ is vanishing while $\eta_1$ is not. As the scalar potential asymptotes to a constant in this case, \eqref{eta-rel} leads to a relation between the deviation from scale invariance and the order of $N$ in $r$:
 \begin{align}
  {\rm \bf Class~I}: \qquad n_s \simeq 1 + \frac{2 \eta_1}{N} \,, \qquad r \sim \frac{1}{N^{-2 \eta_1}} \,. \label{ClassI}
 \end{align}
In order to comply with the assumption that $\epsilon_1 = 0$ one must restrict oneself to $\eta_1 < - \nicefrac12$; in other words, the tensor-to-scalar ratio falls off with more than $1/N$. Thus class I will generically have an $r$ of sub-percentage level. Due to the Lyth bound \cite{Lyth} this corresponds to a possibly (but not necessarily) sub-Planckian field range of the inflaton (more advanced advanced analyses of and counterexamples to this bound can be found in e.g.~\cite{Hotchkiss, Baumann, Boubekeur}). 

For the second possibility we take both $\epsilon_1$ and $\eta_1$ to be non-vanishing. As both asymptotic relations apply, we find a linear relation between these leading coefficients: $\eta_1 = 2 \epsilon_1 - \nicefrac12$. This yields the following cosmological observables:
 \begin{align}
  {\rm \bf Class~II}: \qquad n_s \simeq 1 - \frac{2 \epsilon_1 + 1}{N} \,, \qquad r \simeq \frac{16 \epsilon_1}{N} \,.  \label{ClassII}
 \end{align}
The parameter $\epsilon_1$ is always positive and cannot equal $\nicefrac14$ in this class. In contrast to the previous case, this class has a $1/N$ scaling behaviour of the tensor-to-scalar ratio. For a number of e-folds around 55 this naturally leads to a number of several percents. As implied by the Lyth bound, generic examples of this class are therefore large field inflationary models.

Finally, the third possibility assumes $\eta_1$ to be vanishing while $\epsilon_1$ does not. In this case the asymptotic relations imply $\epsilon_1 = \nicefrac14$ and $\eta_1 = 0$, which was exactly the case that was excluded in the previous analysis. Together these two possibilities therefore give rise to \eqref{ClassII} for all non-negative values of $\epsilon_1$.

Assuming the validity of the asymptotic expansions \eqref{expansion}, we claim that all single-field slow-roll inflationary models fall in either of the two universality classes; the leading contributions of any such models should be of the form \eqref{ClassI} or \eqref{ClassII}, with very specific relations between the expansions of the cosmological parameters. Remarkably, the spectral index cannot be flat but always has at least a $-1/N$ deviation. Without exception, a large number of models indeed satisfy this classification. All $1/N$ expansions given in the encyclopedic survey of inflationary models \cite{ASPIC} are of one of these three forms. Subleading terms will differ between different models, but it will be  observationally difficult to  distinguish between these. 

\begin{figure}[t!]
\centering
\includegraphics[scale=.65]{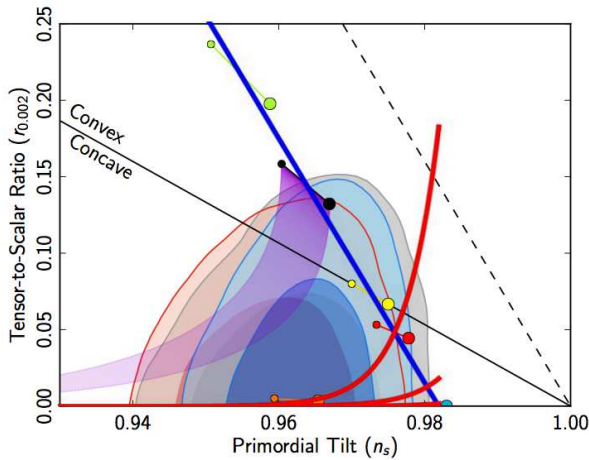}
\caption{\it \small{The inflationary predictions of classes I (thick red) and II (thick blue) with $N=55$ superimposed on Planck data. The three red lines have proportionality constants in $r$ of $(0.1,1,10)$.}}
\label{fig:ClassI}
\vspace{-.3cm}
\end{figure} 

In figure 1 we have superimposed the predictions of both classes with Planck observations. How well do the two universality classes agree with observations, and which parameter values are preferred? A first, rough estimate follows from the spectral index \eqref{ns}, restricting the parameter of class I to 
 \begin{align}
  {\rm Class~I:} \qquad \begin{cases} \eta_1 = - 1.0 \pm 0.2 \,, \quad (N=50) \,, \\ \eta_1 = - 1.2 \pm 0.2 \,, \quad (N=60) \,. \end{cases}
 \end{align}
This implies that the tensor-to-scalar ratio falls off comparable to $1/N^2$; assuming order-$1$ and even order-$10$ coefficients for $r$, this implies an unobservably small value: $r < 0.004$. Instead, the parameter of class II is required to be
  \begin{align}
  {\rm Class~II:} \qquad \begin{cases} \epsilon_1 = 0.5 \pm 0.2 \,, \quad (N=50) \,, \\ \epsilon_1 = 0.7 \pm 0.2 \,, \quad (N=60) \,. \end{cases}
 \end{align}
This translates into the following tensor-to-scalar ratios:
  \begin{align}
  {\rm Class~II:} \qquad \begin{cases} r = 0.16 \pm 0.06 \,, \quad (N=50) \,, \\ r = 0.18 \pm 0.06 \,, \quad (N=60) \,. \end{cases}
 \end{align}
However, as can be seen from figure 1, the actual best fit data will have a lower value of $r$ due to the specific form of class II, and will be close to $\epsilon = \nicefrac14$ and $r \approx 0.07$.  In contrast to most of class II, this model lies within the 95\% confidence level \cite{Planck}. Improved measurements from Planck have the potential to further decrease the error bar, and possibly either detect or restrict to $r < 0.05$. Hence it might well be observationally possible to distinguish between classes I and II. Note that also improvements to the measurement of $n_s$ can play an important role in this endeavour. In particular, a further redshift would strengthen the case for class I, while a blueshift relative to the value \eqref{ns} would give more room for class II and hence large field inflation.

As a final remark, the asymptotic relations only involve the leading order expansion in the inflaton field $\phi$, these will not distinguish between potential energies that are related by a rescaling of the field. In other words, models with $V(\phi)$ and $V(a \phi)$ will lie in the same universality class for an arbitrary real constant $a$. Higher-order terms will depend on $a$, as demonstrated later.

\bigskip

\noindent {\bf Class I examples}
We will now discuss a number of  inflationary models that fall in class I. An important set of small-field models is formed by hilltop inflation \cite{hilltop1, hilltop2}, which has a scalar potential of the form
   \begin{align}
   V = V_0 ( 1 - (\phi / \mu)^n)  \,.
  \end{align}
Indeed it satisfies \eqref{ClassI} with $ \eta_1 = - (n-1)/(n-2)$ for $n>2$. Different polynomials give rise to different scaling behaviours, all of which in the range $\eta_1 < -1$. This set of models therefore populate the parameter space of class I. 

Other models only give rise to specific values of the parameter. An important example is the $R + R^2$ model due to Starobinsky \cite{Starobinsky}, which has also been derived in different contexts recently.  Formulated as a single-field model, the potential energy reads
 \begin{align}
  V = V_0 (1 - e^{\sqrt{2/3} \phi})^2 \,.
 \end{align}
At lowest order it satisfies the criterion of class I with $\eta_1 = -1$, as preferred by Planck. 

Another class of models that was recently proposed on the basis of conformal symmetry arguments are the so-called T-models with \cite{conformal-KL}
 \begin{align}
  V = V_0 \tanh(\phi / \sqrt{6})^{2n} \,. \label{T-model}
 \end{align}
For all values of $n$ this class of models satisfies the class I criterion with $\eta_1 = -1$. Interestingly, for $n=\nicefrac12$ the scaling relation \eqref{eta-rel} is satisfied identically, not only at leading order. This model can therefore be seen as a prototype of class I models with $\eta_1 = -1$. We have not been able to derive similar prototypes for other values of the class I parameter.

A very interesting possibility is new Higgs inflation, where the Standard Model potential has been augmented with a non-minimal coupling $\xi$ to the Ricci scalar \cite{newHiggs}. Formulated in the Einstein frame this model has a potential energy
 \begin{align}
   V = V_0 \frac{\xi^2 \phi^4}{(1 + \xi \phi^2)^2} \,,
 \end{align}
while the kinetic terms are non-canonical and read
 \begin{align}
   \sqrt{-g} [ - \frac{1 + (\xi + 6 \xi^2) \phi^2}{2 (1+ \xi \phi^2)^2} (\partial \phi)^2 ] \,.
 \end{align}
An explicit expression for $n_s$ and $r$ in terms of $N$ is hard to derive, but an accurate approximation was proposed in \cite{Lightinflaton}. Again the expansion of this approximation for non-zero $\xi$ is a class I model with $\eta_1 = -1$. 

Less-known examples leading to different parameter values include arctan inflation with $\eta_1 = - \nicefrac23$, radion gauge inflation with $\eta_1 = - \nicefrac34$ and MSSM inflation with $\eta_1 = -2$. Details of the analysis of these models can be found in \cite{ASPIC}.

\bigskip

\noindent {\bf Class II examples}
We now turn to the generically large-field examples of class II. The first set of models to be considered is chaotic inflation \cite{chaotic}, with a monomial scalar potential:
 \begin{align}
  V = M^4 (\phi / \mu)^{2n} \,,
 \end{align}
with $M, \mu$ constant parameters. The predictions of this model fall in class II with $\epsilon_1 = \nicefrac{n}2$. Thus we also have a set of models that fill out the entire parameter space of class II. Moreover, this set of models can be seen as prototypes of this entire class, as the asymptotic relations \eqref{eta-rel} and \eqref{eps-rel} are satisfied identically and not only at leading order. Of this set, the linear case with the singular values $\epsilon = \nicefrac14$ and $\eta_1 = 0$ seems to be observationally preferred.

A modification of this set of models is the Mexican hat potential,
 \begin{align}
  V = M^4 ( (\phi/\mu)^2 -1)^2 \,,
 \end{align}
also referred to as double well inflation. Inflation takes place between the two minima. In the limit of a super-Planckian vacuum expectation value, $\mu \gg 1$, this leads to another class II model with $\epsilon_1 = \nicefrac12$.

A third example of class II is provided by loop inflation, where the inflationary regime is dominated by radiative corrections:
 \begin{align}
  V = M^4 ( 1 + \alpha \log(\phi)) \,.
 \end{align}
In the limit that the parameter $\alpha$ is vanishingly small, this model allows for an expansion corresponding to class II with $\epsilon_1 = \alpha / 2^{10}$ \cite{ASPIC}.

\bigskip

\noindent {\bf Subleading corrections} We now turn to corrections to the leading order behaviour of a number of models, and find that the asymptotic relations hold beyond leading order.

Starting with hill-top inflation, we have checked in a large number of cases that the asymptotic relations \eqref{eta-rel} and \eqref{eps-sing-rel} even hold beyond leading order. An example is $n=4$, which has the following expansions:
 \begin{align}
  n_s & = 1-\frac{3}{N}+\frac{3 \sqrt{36+\mu^4}}{4 N^2} \,, \notag \\
  r & = \frac{\mu^4}{4 N^3}-\frac{3 \mu^4 \sqrt{36+\mu^4}}{16 N^4} \,, \notag \\
 V & = V_0 ( 1-\frac{\mu^4}{64 N^2}+\frac{\mu^4\sqrt{36+\mu^4}}{128 N^3}) \,. \label{hilltop-c}
 \end{align}
From these it follows that the asymptotic relation \eqref{eta-rel} holds both at the leading order, as discussed before, but also at next-to-leading order: the $1/N$ contributions cancel on both sides. Thus both the leading scaling behaviour, as well as the first correction this, agree. At order $1/N^2$ there are contributions to both sides, but the normalised coefficients do not agree. Similarly, the asymptotic relation \eqref{eps-sing-rel} for the singular case of $\epsilon_1 = 0$ agrees at the two lowest orders $1$ and $1/N^2$, up to a single overall constant.

Next we turn to the Starobinsky model. In fact we will analyse a generalisation thereof that was recently proposed in the context of non-minimally coupled models \cite{non-minimal-KL} with
 \begin{align}
 V = V_0 (1- e^{- \sqrt{2/3} \phi})^{2n} \,,
 \end{align}
have the following next-to-leading order expansion:
 \begin{align}
  n_s & = 1 - \frac2N + \frac{3}{2n} \frac{\log(N)}{N^2} \,, \notag \\
  r & = \frac{12}{N^2} - \frac{18}{n} \frac{\log(N)}{N^3} \,, \notag \\
  V & = V_0 ( 1 - \frac32 \frac1N + \frac98 \frac{\log(N)}{N^2} ) \,. \label{Star1-c}
 \end{align}
Again the asymptotic relation \eqref{eta-rel} is satisfied at lowest order and the first correction, the latter being the absence of $\log(N) / N$ terms. At higher orders, e.g.~$1/N$, the two sides start to deviate. The other asymptotic relation \eqref{eps-sing-rel} agrees at order $1$ and $1/N$. A similar generalisation is
 \begin{align}
  V = V_0 (1- e^{- \sqrt{2n/3} \phi})^2 \,,
 \end{align}
where again the original corresponds to $n = 1$, while $n=3$ was recently proposed in a supergravity context \cite{RSZ}. This set of models has a very comparable expansion:
 \begin{align}
  n_s & = 1-\frac{2}{N}+ \frac{3}{2n} \frac{\log(N)}{N^2} \,, \notag \displaybreak[2] \\
 r & =\frac{12}{n N^2} - \frac{18}{n^2} \frac{\log(N)}{N^3} \,, \notag \\
  V & = V_0 (1- \frac{3}{2 n N} +\frac{9}{8n^2} \frac{\log(N)}{N^2}) \,. \label{Star2-c} 
\end{align}
The two large-$N$ relations agree in the same qualitative manner as the previous model.

Finally, the leading order of the T-models \eqref{T-model} agree with that of Starobinsky, being class I with $\eta_1 = -1$, while the corrections are
 \begin{align}
  n_s & = 1-\frac{2}{N}+\frac{-3 n+\sqrt{9+12 n^2}}{2 n N^2} \,, \notag \\
  r & = \frac{12}{N^2}-\frac{6 \sqrt{9+12 n^2}}{n N^3} \,, \notag \\
 V & = V_0 ( 1-\frac{3}{2 N}+\frac{9 n+3 \sqrt{9+12 n^2}}{8 n N^2} ) \,. \label{T-model-c}
 \end{align}
Again the large-$N$ relation \eqref{eta-rel} holds both at leading as well as next-to-leading order; both sides have terms that scale as $1$ and $1/N$ and, moreover, the ratio between the coefficients is identical on both sides. The latter ceases to be true at higher order for generic $n$. For $n= \nicefrac12$ this model is the prototype of its class and hence satisfies the scaling relation at all orders. The other asymptotic relation \eqref{eps-sing-rel} holds at lowest and next-to-lowest order in this case, again being $1$ and $1/N$. Somewhat surprisingly, higher-order terms of this equation always differ, even for the special case $n = \nicefrac12$. 

\bigskip

\noindent {\bf Discussion}
We have demonstrated that all single-field, slow-roll inflationary models  (without dissipative effects as \cite{warminflation}) whose slow-roll parameters scale with $1/N$ or a higher power thereof reduce to either of the one-parameter universality classes \eqref{ClassI} and \eqref{ClassII} in the large-$N$ limit. Subleading corrections, that are found to satisfy the same asymptotic relations, will be model-dependent but unobservably small. In conjunction with the value of the spectral index as measured by Planck, this leads to either an unobservably small tensor-to-scalar ratio $r < 0.01$ (class I) or a  value around $r \approx 0.07$ (class II). The latter is rather close to the observational limit of $r<0.11$. Excitingly, it could even be detected or ruled out by improved data from Planck, which might reduce the error bar in $r$ to $0.05$. The distinction between class I and II and, as a consequence, the $N$-dependence of $r$ could thus be observationally settled. As stressed before, on account of the relations between $n_s$ and $r$, this issue is also highly sensitive to improved measuments of the spectral index.

This paper builds on a perturbative expansion in $1/N$ as a naturally small number. An additional small parameter could complicate the above analysis. For instance, it could  invalidate the order-of-magnitude reasoning that leads to an unobservably small $r$ in class I. The scaling behaviour \eqref{ClassI} would still hold but the  proportionality constant could be very large, rendering $r$ super-percent level (an example would be new Higgs inflation with a very small coupling $\xi$). Such models would require an additional argument for the smallness of the extra parameter, however. Similarly, the assumption of a leading power-law dependence in $1/N$ could be violated. An example is natural inflation \cite{natural}, indicated by the shady region in figure 1, whose deviation from scale invariance is a constant plus non-perturbative terms in $1/N$. Another model, termed \Kahler modulus inflation in \cite{ASPIC}, has $\log(N)/N$ terms at leading order. However, in order to comply with Planck, this type of models generically also requires small parameters.The analysis presented here therefore covers an important set of inflationary models that are naturally viable.

\bigskip

\noindent {\bf Acknowledgments} 
We acknowledge stimulating discussions with Andrea Borghese, Sander Mooij, Marieke Postma, Marco Scalisi and Ivonne Zavala and financial support by a VIDI grant from NWO.

\providecommand{\href}[2]{#2}\begingroup\raggedright\endgroup

\end{document}